\documentclass[aps,prb,twocolumn,groupedaddress,showpacs]{revtex4-1}

\usepackage{graphicx}
\usepackage{color}
\usepackage{ulem}
\usepackage{verbatim}
\usepackage{hyperref}

\begin{document}

\title{Quasi-2D behavior of 112-type iron-based superconductors}

\author{D. S\'o\~nora$^1$}
\author{C. Carballeira$^1$}
\author{J.J. Ponte$^2$}
\author{Tao Xie$^{3,5}$}
\author{Huiqian Luo$^3$}
\author{Shiliang Li$^{3,4,5}$}
\author{J. Mosqueira$^1$}
\email[]{j.mosqueira@usc.es}

\affiliation{$^1$LBTS, Departamento de F\'isica de Part\'iculas, Universidade de Santiago de Compostela, E-15782 Santiago de Compostela, Spain}

\affiliation{$^2$Unidade de Magnetosusceptibilidade, RIAIDT, Universidade de Santiago de Compostela, E-15782 Santiago de Compostela, Spain}

\affiliation{$^3$Beijing National Laboratory for Condensed Matter Physics, Institute of Physics, Chinese Academy of Sciences, Beijing 100190, People's Republic of China}

\affiliation{$^4$Collaborative Innovation Center of Quantum Matter, Beijing 100190, China}

\affiliation{$^5$University of Chinese Academy of Sciences, Beijing 100190, China}

\date{\today}

\begin{abstract}
Fluctuation magnetoconductivity and magnetization above the superconducting transition temperature ($T_c$) are measured on the recently discovered 112 family of iron-based superconductors (IBS), Ca$_{1-x}$La$_x$Fe$_{1-y}$Ni$_y$As$_2$, which presents an extra As-As chain spacer-layer. The analysis in terms of a generalization of the Lawrence-Doniach (LD) approach to finite applied magnetic fields indicates that these compounds are among the most anisotropic IBS ($\gamma$ up to $\sim30$), and provides a compelling evidence of a \textit{quasi}-two-dimensional behavior for doping levels near the optimal one. 
\end{abstract}


\maketitle

\section{Introduction}

All families of iron-based superconductors (IBS) share a similar crystal structure consisting of FeAs superconducting layers separated by spacer layers that determine many of their properties.\cite{[{For a recent review see, e.g., }]reviews} Recently it has been discovered a new class of IBS (the 112 family) based in the compound Ca$_{1-x}$La$_x$FeAs$_2$,\cite{Katayama13} that has raised a great interest.\cite{Kudo14,Kudo14_2,Kawasaki15,Li15,Wu15topoSC,Jiang16_2,Jiang16,Liu16Dirac} In addition to the Ca/La spacer layer, these compounds present an extra spacer layer with zigzag As chains that introduces an additional electron band near the Fermi level.\cite{Li15,Jiang16} In agreement with a previous theoretical study\cite{Wu15topoSC} this band presents a Dirac-cone structure,\cite{Liu16Dirac} which led to the recent proposal that these compounds may behave below $T_c$ as natural topological superconductors.\cite{Wu15topoSC,Liu16Dirac} The extra As layer also increases significantly the distance between the superconducting FeAs layers (up to $\sim10.3$~\r{A}) as compared with the most studied IBS families. This could strongly enhance the superconducting anisotropy, and even affect the spatial dimensionality of the superconducting order parameter, at present an open issue in IBS. For instance, compounds with smaller FeAs layers interdistance were claimed to present 2D characteristics (e.g., LiFeAs,\cite{Song12,Rullier12} FeSe$_{1-x}$Te$_x$,\cite{Pandya10} and SmFeAsO \cite{Pallecchi09,Liu10_1,*Liu10_2}), although recent works in the same or similar compounds suggest a 3D anisotropic behavior.\cite{Putti10,Welp11,Liu11SSC,Pandya11,comment,Ahmad14,Ahmad16}

Here we study the anisotropy and dimensionality of high-quality 112 single crystals through measurements of the conductivity induced by superconducting fluctuations above $T_c$, $\Delta\sigma$. Fluctuation effects are also a powerful tool to determine other fundamental superconducting parameters as the coherence lengths or the critical fields,\cite{tinkham,bookLarkin} and are even sensitive to the multiband electronic structure.\cite{koshelev,ramos15,ikeda} The experiments were performed with magnetic fields $H$ up to 9~T applied both parallel and perpendicular to the FeAs ($ab$) layers. These field amplitudes are large enough to explore the so-called \textit{finite-field} or \textit{Prange} fluctuation regime,\cite{tinkham,bookLarkin} and to quench the unconventional behavior observed in IBS below $\sim 1$~T, usually attributed to phase fluctuations\cite{phasefluc_1,*phasefluc_2} or to a $T_c$ distribution.\cite{rey13,ramos15} To analyze the data the Gaussian LD approach for $\Delta\sigma$ (see Ref.~\onlinecite{Hikami88}) is generalized here to the finite-field regime and to high reduced-temperatures through the introduction of a total-energy cutoff.\cite{Vidal02} These data are complemented with measurements in another single crystal of the fluctuation-induced magnetization around $T_c$, $\Delta M$. This observable is proportional to the effective superconducting volume fraction and confirms the bulk nature of the superconductivity in these materials. It also provides an important consistency check of the results.

Details of the crystals growth and characterization are presented in \S~II, the measurements and analysis of $\Delta\sigma$ and $\Delta M$ in \S~III and \S~IV, respectively, the discussion of the results in \S~V, and the conclusions in \S~VI. 

\section{Crystal growth and characterization}

The composition of the single crystals used in the experiments is \mbox{Ca$_{1-x}$La$_x$Fe$_{1-y}$Ni$_{y}$As$_2$} with $x=0.17-0.20$ and $y=0.044(3)$. The partial substitution of Fe by Ni (or Co) improves the superconducting properties and sharpens the superconducting transition,\cite{Jiang16_2,Yakita15} which is essential to study critical phenomena around $T_c$. They were grown by a self-flux method. The precursor materials CaAs, LaAs, FeAs, and NiAs were grinded with a molar ratio $3.7:0.3:0.95:0.05$. The mixed powder was then pressed into a pellet, loaded into an Al$_2$O$_3$ crucible and sealed into a quartz tube. The ampoule was heated to 1180$^\circ$C, slowly cooled down to 950$^\circ$C, and then to room temperature. After cracking the melted pellet, shining plate-like single crystals with typical size $1\times1\times0.05$~mm$^3$ could be obtained. A thorough description may be seen in Ref.~\onlinecite{Luo17}.

\begin{figure}[t]
\begin{center}
\includegraphics[scale=.5]{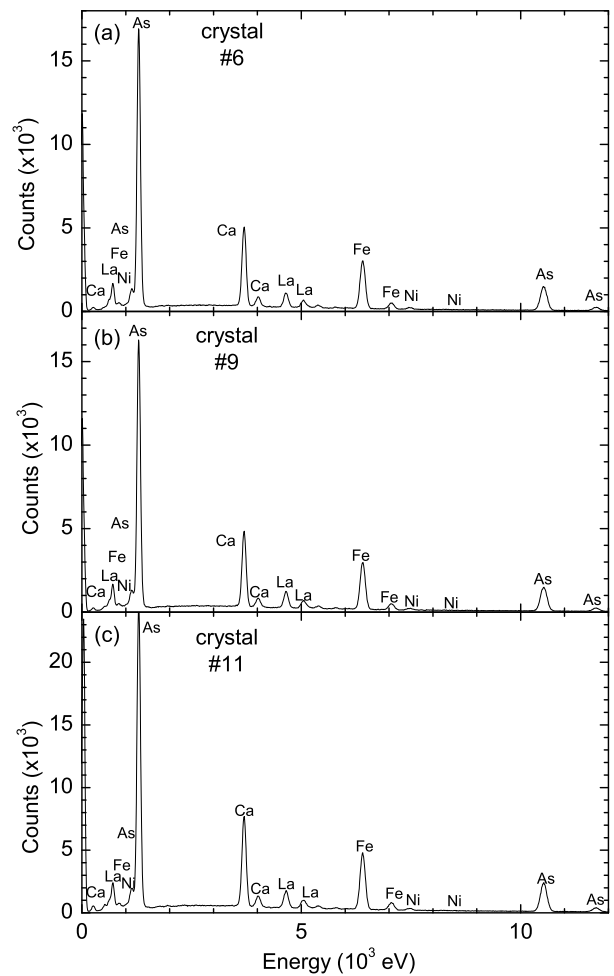}
\caption{Examples of EDX spectrum measured in the studied crystals. }
\label{FigEDX}
\end{center}
\end{figure}

The stoichiometry of the three crystals used in the experiments (\#6, \#9 and \#11) was checked by energy-dispersive x-ray spectroscopy (EDX), performed with a Zeiss FE-SEM Ultra Plus system. EDX spectra were taken in five different points in each crystal (some examples are presented in Fig.~\ref{FigEDX}). The average stoichiometry is presented in Table~I, where the number in parenthesis represent the standard deviation. The differences from crystal to crystal in the average La content are slightly beyond the deviation, which will be useful to explore the dependence of superconducting parameters on the La doping level.

\begin{figure}[t]
\begin{center}
\includegraphics[scale=.5]{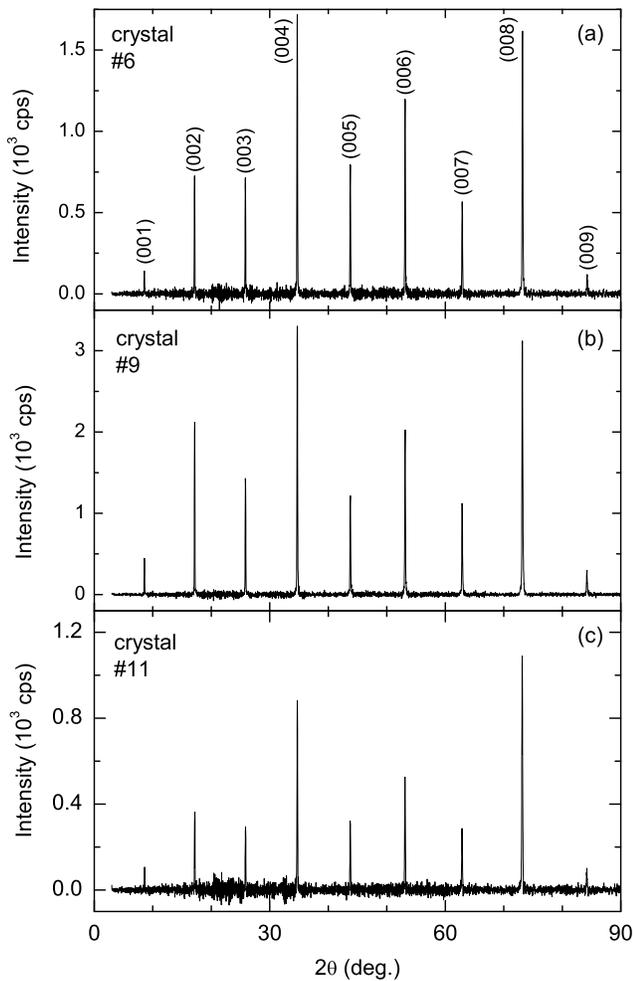}
\caption{XRD pattern of the studied crystals (already corrected for a background contribution).}
\label{FigRX}
\end{center}
\end{figure}

The crystallographic structure was studied by x-ray diffraction (XRD) by using a Rigaku MiniFlex~II diffractometer with a Cu-target. The $\theta-2\theta$ patterns (see Fig.~\ref{FigRX}) present only $(00l)$ reflections, which indicates the excellent structural quality of the crystals. The resulting $c$-axis lattice parameter (that is the same as the FeAs layers interdistance, $s$) is about $10.34$~\r{A} (see Table I), in agreement with data in the literature for crystals with a similar composition.\cite{Katayama13} 

\begin{table}[ht]
\caption{Average composition and interlayer distance of the studied samples, as follows from EDX and XRD.}
\begin{ruledtabular}
\begin{tabular}{l|ccccc|c}
 Crys-&   \multicolumn{5}{c|}{Composition}  & $s$ \\
tal & Ca&La&Fe&Ni&As & (\r{A}) \\
\hline
\#6 & 0.829(5) & 0.172(2) & 0.925(8) & 0.044(3) & 2.030(7) & 10.336(1) \\
\#9 & 0.802(4) & 0.199(7) & 0.921(3) & 0.044(2) & 2.034(6) & 10.343(1) \\
\#11 & 0.833(5) & 0.176(3) & 0.950(7) & 0.045(3) & 1.996(7) & 10.348(1) \\
\end{tabular}
\end{ruledtabular}
\end{table}

\section{Paraconductivity and magnetoconductivity induced by fluctuations}

\subsection{Experimental details and results}

\begin{figure}[t]
\begin{center}
\includegraphics[scale=.45]{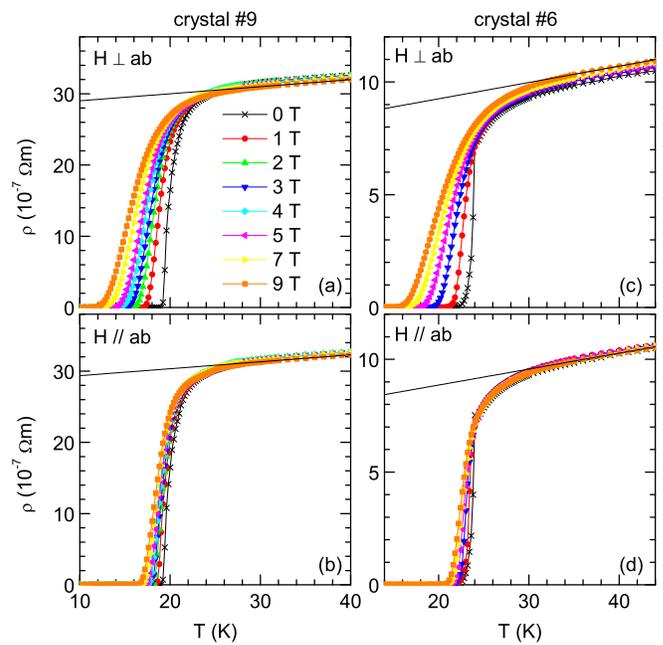}
\caption{(Color online) $T$-dependence of the resistivity around $T_c$ of the studied crystals for both field orientations. The lines are examples (for $\mu_0H=9$~T) of the background contribution, as determined by a linear fit above 35~K, where fluctuation effects are expected to be negligible.}
\label{Figrho}
\end{center}
\end{figure}

The resistivity along the \textit{ab} layers, $\rho$, was measured in crystals \#6 and \#9 with a Quantum Design's Physical Property Measurement System (PPMS) by using a four-wire technique with 1~mA excitation current at 71 Hz. The $\rho(T)_H$ behavior around $T_c$ for both $H\parallel ab$ and $H\perp ab$ is presented in Fig.~\ref{Figrho}. The $T_c$ values (see Table II) were estimated from the midpoint of the resistive transition in absence of field. The slight difference may be attributed to the above mentioned differences in the La content. The transition \textit{half-widths}, estimated from the 50\%-10\% criterion (above 50\% intrinsic fluctuation effects also contribute to the transition widening), are around 0.6~K.  This allowed to investigate fluctuation effects down to reduced temperatures $\varepsilon\equiv\ln(T/T_c)$ as low as 0.03. The resistivity rounding due to fluctuations extends in both samples up to $\sim30$~K ($\varepsilon\approx0.4$), and is larger in amplitude than in other IBS with a similar $T_c$ and normal-state resistivity.\cite{rey13,rey14} The Aslamazov-Larkin (AL) model for 3D anisotropic superconductors predicts that $\Delta\sigma\propto\xi_c^{-1}(0)$,\cite{tinkham,bookLarkin} where $\xi_c(0)$ is the $c$-axis coherence length amplitude. Thus, the enhanced fluctuation effects in 112-IBS is a first indication that these materials present a smaller $\xi_c(0)$, and may present a quasi-2D behavior if it is smaller than the FeAs-layers distance. This seems to be the case in view of the almost inappreciable $T_c(H)$ shift for $H\parallel ab$, mainly in crystal \#6.  

\begin{table}[ht]
\caption{Superconducting parameters of the studied crystals obtained from the analysis of fluctuation effects in the indicated observable.}
\begin{ruledtabular}
\begin{tabular}{ccccccc}
Crystal & obs. & $T_c$ & $\xi_c(0)$ & $\xi_{ab}(0)$ & $r$ & $\gamma$\\
 & & (K) & (\r{A}) & (\r{A}) & & \\
\hline
\#6 & $\Delta\sigma$ & 23.9 & 0.65 & 19.3 & 0.016 & 29.7\\
\#9 & $\Delta\sigma$ & 19.9 & 3.8 & 32.7 & 0.55 & 8.5 \\
\#11 & $\Delta M$    & 21.8 & 1.9 & 27.0 & 0.13 & 14\\
\end{tabular}
\end{ruledtabular}
\end{table}

\subsection{Analysis of fluctuation effects above $T_c$}

The fluctuation contribution to the conductivity $\Delta\sigma(T)_H$ was obtained from $\rho(T)_H$ through $\Delta\sigma(T)_H=1/\rho(T)_H-1/\rho_B(T)_H$, where the background resistivity $\rho_B(T)_H$ was obtained for each field by a linear fit from 35 to 40~K, above the onset of fluctuation effects. The upper limit was chosen to avoid a subtle change in the $\rho(T)_H$ behavior at higher temperatures, qualitatively similar to the one observed well above $T_c$ in other 112 compounds and attributed to magnetic/structural phase transitions.\cite{Jiang16_2} 

\subsubsection{Crystal \#9}

\begin{figure}[t]
\begin{center}
\includegraphics[scale=.5]{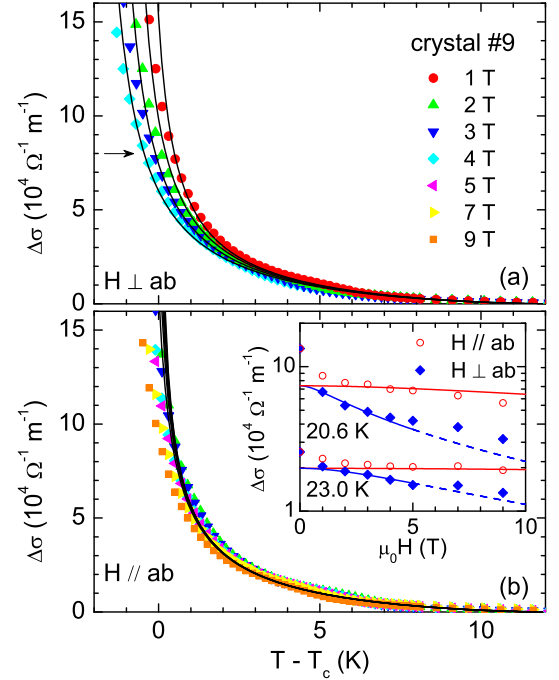}
\caption{(Color online) Analysis of the results for crystal \#9. (a) $T$-dependence of $\Delta\sigma$ for $H\perp ab$. The lines are the best fit of Eq.~(\ref{3DH}) to the data up to the $\Delta\sigma$ value indicated by the arrow. (b) $T$-dependence of $\Delta\sigma$ for $H\parallel ab$. Inset: $H$-dependence for different temperatures above $T_c$. The lines in the main panel and in the inset were evaluated by using in Eq.~(\ref{3DH}) the superconducting parameters obtained in (a). See the main text for details.}
\label{Figsigma9}
\end{center}
\end{figure}

The resulting $\Delta\sigma(T)_H$ for this crystal is presented in Fig.~\ref{Figsigma9}. These data are first analyzed in terms of the Gaussian 3D-anisotropic Ginzburg-Landau (GL) approach developed in Ref.~\onlinecite{rey13}, that includes a cutoff in the energy of the fluctuation modes,\cite{Vidal02} and is valid beyond the zero-field limit,
\begin{eqnarray}
&&\Delta\sigma_{\rm 3D}=\frac{e^2}{32\hbar \pi\xi_c(0)}\sqrt{\frac{2}{h}}\int_0^{\sqrt{\frac{c-\varepsilon}{2h}}}{\rm d}x
\left[\psi^1\left(\frac{\varepsilon+h}{2h}+x^2\right)-\right.\nonumber\\
&&\left.\psi^1\left(\frac{c+h}{2h}+x^2\right)\right].
\label{3DH}
\end{eqnarray}
Here $\psi^1$ is the first derivative of the digamma function, $e$ is the electron charge, $\hbar$ is the reduced Planck constant, $h=H/H_{c2}(0)$ is the reduced magnetic field, $H_{c2}(0)$ is the upper critical field linearly extrapolated to $T=0$~K ($H_{c2}^c$ when $H\perp ab$, and $H_{c2}^{ab}$ when $H//ab$), and $c$ is the cutoff constant,\cite{Vidal02} that corresponds to the $\varepsilon$-value for the onset of fluctuation effects. As in this crystal $\Delta\sigma$ is found to vanish at $T_{\rm onset}-T_c\approx12$~K, we approximated $c=\ln(T_{\rm onset}/T_c)\approx0.47$. Eq.~(\ref{3DH}) is valid up to reduced magnetic fields of the order of $h\sim c/2\approx0.2$ (see Ref.~\onlinecite{rey13}). As expected, in the zero-field limit ($h\ll \varepsilon$) and in absence of cutoff ($c\to\infty$) it reduces to the conventional 3D-AL expression.\cite{rey13}

The analysis for $H\perp ab$ is presented in Fig.~\ref{Figsigma9}(a). The lines are the best fit of Eq.~(\ref{3DH}) to the data obtained under magnetic fields from 1 to 4~T and up to $\Delta\sigma=8\times10^4$~$(\Omega$m)$^{-1}$ (indicated by an arrow). We have checked that extending the fitting region above this $\Delta\sigma$ value increases significantly the root-mean-square deviation (RMSD). Then, this $\Delta\sigma$ limit may be associated to the onset of the critical region, where fluctuation effects are so large that the Gaussian approximation is no longer valid and Eq.~(\ref{3DH}) is not applicable. In what concerns the magnetic field range, data obtained with $\mu_0H\geq5$~T were excluded because they considerably worsened the fit quality. In view of the $H_{c2}^c(0)$ value resulting from the analysis (see below), this may be associated to the limit of applicability of the theory. In turn, we excluded data below 1~T because in this region $\Delta\sigma(H)$ presents an anomalous upturn [see the inset in Fig.~\ref{Figsigma9}(b)], an effect already observed in other IBS and attributed to the possible presence of phase fluctuations\cite{phasefluc_1,*phasefluc_2} but also to a $T_c$ distribution.\cite{rey13,ramos15} 

The values obtained for the two fitting parameters are $\xi_c(0)=3.8$~\r{A} and $\mu_0H_{c2}^c(0)=30.7$~T, which leads to an in-plane coherence length amplitude of $\xi_{ab}(0)=\sqrt{\phi_0/2\pi\mu_0H_{c2}^c(0)}=32.7$~\r{A}. The corresponding anisotropy factor $\gamma=\xi_{ab}(0)/\xi_c(0)$ is as large as 8.5, but still consistent with the 3D behavior because the LD parameter $r\equiv[2\xi_c(0)/s]^2$, which is associated to the reduced temperature for the 3D-2D crossover,\cite{tinkham} is $\sim0.55$, above the onset of fluctuation effects. 

The analysis for $H\parallel ab$ is presented in Fig.~\ref{Figsigma9}(b), where the solid lines were obtained without free parameters, by using in Eq.~(\ref{3DH}) the above $\xi_c(0)$ value and $\mu_0H_{c2}^{ab}(0)=\gamma \mu_0H_{c2}^c(0)=261$~T. The agreement with the experimental data is also excellent, which is an important consistency check of our results. For completeness, in the inset of Fig.~4(b) the $H$ dependence of $\Delta\sigma$ is presented for both field orientations and for two temperatures above $T_c$. The lines were obtained by using in Eq.~(\ref{3DH}) the above superconducting parameters. The dashed line corresponds to $H\stackrel{>}{_\sim}0.2H_{c2}^c(0)$, where the theory is no longer applicable.  

\subsubsection{Crystal \#6}

\begin{figure}[t]
\begin{center}
\includegraphics[scale=.5]{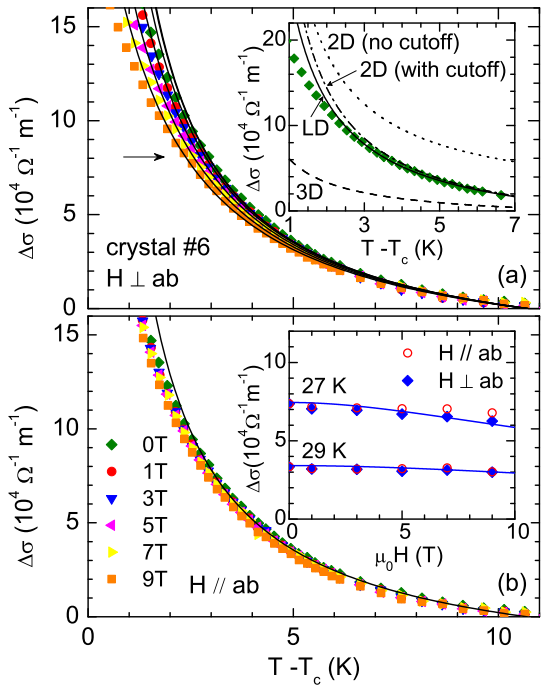}
\caption{(Color online) Analysis of the results for crystal \#6. (a) $T$-dependence of $\Delta\sigma$ for $H\perp ab$. The lines are the best fit of Eq.~(\ref{LDH}) to the data up to the $\Delta\sigma$ value indicated by the arrow. Inset: Comparison of the $H=0$ data with the 2D, 3D and LD approaches (see the main text for details). (b) $T$-dependence for $H\parallel ab$. The line was evaluated with the zero-field LD expression, Eq.~(\ref{LDH0}), by using the $r$ value obtained in (a). Inset: $H$-dependence of $\Delta\sigma$ above $T_c$ for both field orientations. The data for $H\parallel ab$ are almost $H$-independent, as expected for a (quasi)-2D superconductor. The lines correspond to $H\perp ab$ and were evaluated by using in Eq.~(\ref{LDH}) the same parameters as in (a).}
\label{Figsigma6}
\end{center}
\end{figure}

The $\Delta\sigma$ behavior of this crystal is presented in Fig.~\ref{Figsigma6}. As it may be seen in the inset in (b), $\Delta\sigma$ presents a monotonous behavior when $H\to 0$ near $T_c$, suggesting that $T_c$ inhomogeneities or phase fluctuations play a negligible role in this sample. A first comparison with the theory may be then done with the data obtained with $H=0$. As it is shown in the inset in Fig.~\ref{Figsigma6}(a), the $\Delta\sigma(T,H=0)$ amplitude is appreciably larger than the one predicted by the 3D approach by using the $\xi_c(0)$ value previously found in crystal \#9, and $c=0.37$ (according to the $\varepsilon$-value at which $\Delta\sigma$ vanishes in crystal \#6). A smaller $\xi_c(0)$ value (about 1~\r{A}) leads to an acceptable agreement with the data, but is inconsistent with a 3D behavior (it would lead to $r\approx0.04$, so that the system should behave as 2D in almost all the accessible $\varepsilon$ range). In turn, the conventional 2D-AL approach, $\Delta\sigma=e^2/16\hbar s\varepsilon$ (dotted line) where $s$ is the FeAs-layers interdistance, strongly overestimates the experimental $\Delta\sigma$. The agreement improves with the introduction of an energy cutoff, which leads to $\Delta\sigma=e^2/16\hbar s(\varepsilon^{-1}-c^{-1})$ (dot-dashed line, see below), but only at high reduced temperatures. This suggests that a intermediate-dimensionality LD approach is needed. 

A LD expression for $\Delta\sigma$ under finite applied magnetic fields may be obtained by adapting Eq.~(B.18) of Ref.~\onlinecite{rey13} (giving the fluctuation-induced conductivity in 3D as a sum over the contributions of  different Landau levels) to the quasi-2D case by introducing the appropriate  out-of-plane spectrum of the fluctuations\cite{Hikami88} ({\it i.e.}, substituting $\omega_{k_{z}}^{\rm 3D}=\xi_c^2(0)k_{z}^2$ by $\omega_{k_{z}}^{\rm LD}=r[1-\cos(k_{z}s)]/2$) and taking into account the structural cutoff in the $z$-direction through $|k_z|\leq \pi/s$. This leads to
\begin{equation}
\Delta \sigma_{\rm LD}=
\frac{e^2h}{16 \pi \hbar }
\int_{-\pi/s}^{\pi/s}{\rm d}k_{z}
\sum_n [\varepsilon+h(2n+1)+\omega_{k_{z}}^{\rm LD}]^{-2},
\label{summagnetoLD}
\end{equation}
where  the sum over Landau-levels is to be performed up to $n_{\rm max}=(c-\varepsilon)/2h-1$, resulting
\begin{eqnarray}
&&\Delta \sigma_{\rm LD}=\frac{e^2}{64 \pi \hbar}\frac{1}{h}\int_{-\pi/s}^{\pi/s}{\rm d}k_{z}
\left[\psi^{1}\left(\frac{\varepsilon+h+\omega_{k_{z}}^{\rm LD}}{2h}\right)-\right.\nonumber\\
&&\left.\psi^{1}\left(\frac{c+h+\omega_{k_{z}}^{\rm LD}}{2h}\right)
\right].
\label{LDH}
\end{eqnarray}
In the low field limit $h \ll \varepsilon$ this expression reduces to
\begin{equation}
\Delta \sigma_{\rm LD}=\frac{e^2  }{16 \hbar s}
\left[\frac{1}{\sqrt{\varepsilon(\varepsilon+r)}}-\frac{1}{\sqrt{c(c+r)}}\right],
\label{LDH0}
\end{equation}
that in the absence of cutoff ($c\to\infty$) leads to the conventional LD paraconductivity.\cite{bookLarkin}

The solid line in the inset of Fig.~\ref{Figsigma6}(a) is the best fit of Eq.~(\ref{LDH0}) to the data obtained with $H=0$ and up to $\Delta\sigma=8\times10^4$~$(\Omega$m)$^{-1}$. By using a fitting region above this $\Delta\sigma$ value increases significantly the RMSD, so this $\Delta\sigma$ limit may be associated to the onset of the critical region where Eq.~(\ref{LDH0}) is not applicable. The value obtained for the only free parameter is $r=0.016$, which leads to $\xi_c(0)=0.65$~\r{A}, a value more than one order of magnitude smaller than the FeAs layers interdistance. The solid lines in the main panel of Fig.~\ref{Figsigma6}(a) are the best fit of Eq.~(\ref{LDH}) to the data obtained with $H\perp ab$ up to 9~T and up to the same limit, $\Delta\sigma=8\times10^4$~$(\Omega$m)$^{-1}$. In this case we used the above $r$ and $c$ values, and obtained for the only free parameter $\mu_0H_{c2}^c(0)=88.1$~T. This value leads to $\xi_{ab}(0)=[\phi_0/2\pi\mu_0 H_{c2}^c(0)]^{1/2}=19.3$~\r{A}, and to an anisotropy factor as large as $\gamma=\xi_{ab}(0)/\xi_c(0)\approx 29.7$. This result is confirmed by the inappreciable effect on $\Delta\sigma$ of magnetic fields parallel to the $ab$ layers, see Fig.~\ref{Figsigma6}(b). 

\section{Magnetization induced by fluctuations around $T_c$}

\subsection{Experimental details and results}

In order to confirm the above results we have performed additional measurements of the magnetization ($M$) induced by superconducting fluctuations in crystal \#11. As commented above, this observable is proportional to the effective superconducting volume fraction, and is suitable to confirm the bulk nature of the superconductivity in these compounds. The measurements were performed with a Quantum Design's magnetometer (model MPMS-XL). The crystal was measured with $H$ perpendicular to the $ab$ layers. For that we used a quartz sample holder (0.3 cm in diameter, 22 cm in length) with a $\sim0.3$ mm wide groove perpendicular to its axis, into which the crystal was glued with GE varnish. Two plastic rods at the sample holder ends ($\sim0.3$ mm smaller than the sample space diameter) ensured that its alignment was better than $0.1^\circ$. 


\begin{figure}[t]
\begin{center}
\includegraphics[scale=.5]{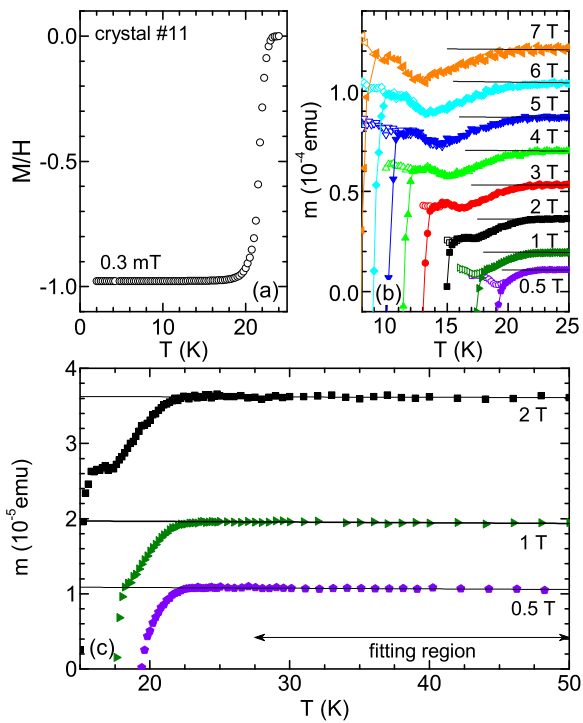}
\caption{(Color online) (a) $T$-dependence of the low-field (0.3 mT) ZFC magnetic susceptibility of crystal \#11 (already corrected for demagnetizing effects). (b) Detail of the $T$-dependence of the as-measured magnetic moment around $T_c$. Solid and open symbols were obtained under ZFC and FC conditions, respectively. The diamagnetism above $T_c$ is unobservable in this scale. (c) Examples of the $T$-dependence up to 50 K, where the normal-state backgrounds (lines) were determined by a linear fit in the indicated region.}
\label{Figraw}
\end{center}
\end{figure}

As a first magnetic characterization, in Fig.~\ref{Figraw}(a) it is presented the temperature dependence of the low field (0.3 mT) zero-field-cooled (ZFC) magnetic susceptibility, $\chi=M/H$. This measurement is corrected for demagnetizing effects by using as demagnetizing factor $D=0.86$, as it results by approximating the crystal shape by an ellipsoid. As it may be seen, $\chi$ is near the ideal shielding value of -1 just below the diamagnetic transition. $T_c\approx21.8$~K was estimated as the temperature at which $d\chi/dT$ is maximum, and the transition width as $\Delta T_c=T(\chi=0)-T_c\approx1$~K, that will allow to study the fluctuation-induced magnetization in a wide temperature region above $T_c$ (see below).

To measure the effect of superconducting fluctuations above $T_c$ (which is in the $10^{-6}$ emu range), for each temperature we averaged eight independent measurements, from which we excluded the ones that deviate more than the standard deviation from the average value. The final resolution in magnetic moment, $m$, was in the $\sim10^{-8}$ emu range. The as-measured $m(T)$ data around $T_c$ are presented in Fig.~\ref{Figraw}(b). The solid (open) data points were obtained under ZFC (FC) conditions. As it is clearly seen, the reversible region extends a few degrees below $T_c$, allowing to study the critical fluctuation regime. Just above the irreversibility temperature $m(T)$ presents an upturn that grows in amplitude with $H$. A very similar effect has also been observed in low-$T_c$ alloys and has been attributed to surface superconductivity.\cite{Gollub73} In the following we will restrict the analysis of fluctuation effects to temperatures above this anomaly. 

Some examples of the $m(T)_H$ behavior above $T_c$ are presented in Fig.~\ref{Figraw}(c). In view of the almost constant temperature dependence, the background magnetic moment was determined by fitting a linear function, $m_B(T)=a+bT$, between 27.5~K (a temperature above which the rounding due to fluctuation effects is not appreciable) and 50~K. The temperature dependence around $T_c$ of the magnetization induced by fluctuations $\Delta M=(m-m_B)/VH$ (where $V$ is the crystal volume) is presented in the inset of Fig.~\ref{Figscal}. 

\subsection{Analysis in the critical region}

\begin{figure}[t]
\begin{center}
\includegraphics[scale=.5]{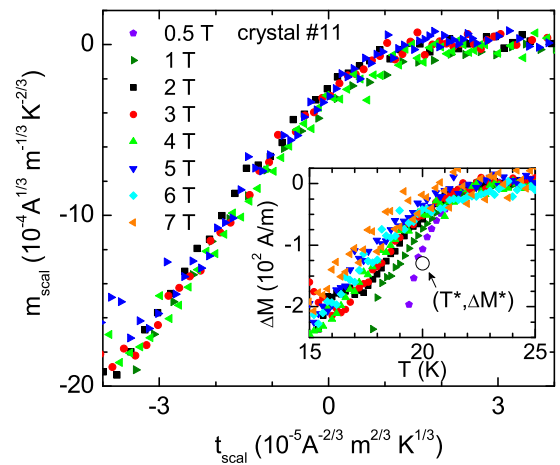}
\caption{(Color online) 3D-GL scaling of the magnetization in the critical region. For clarity the noisy data for 6~T and 7~T were not included in the representation, but they are still consistent with the scaling. Inset: $T$-dependence of the unscaled $\Delta M$ data around $T_c$. The circle is the prediction for the crossing point of 2D superconductors. See the main text for details.}
\label{Figscal}
\end{center}
\end{figure}

A first direct analysis of the data may be done through Te\u{s}anovi\'c's approach for the magnetization in the critical region of 2D materials.\cite{Tesanovic92} This model predicts that the $\Delta M(T)$ curves obtained under different $H$ amplitudes cross at $\Delta M^*=-k_BT^*/\phi_0s$, $\Delta M^*$ and $T^*$ being the crossing point coordinates. In the case of crystal \#11 the crossing occurs at much smaller $\Delta M$ amplitude (the the inset in Fig.~\ref{Figscal}), suggesting that the behavior in the critical region may be closer to the one of a 3D superconductor. In this region, the 3D-GL approach in the lowest-Landau-level approximation predicts a scaling behavior in the variables\cite{Ullah90,Ullah91} 
\begin{equation}
m_{\rm scal}\equiv \frac{\Delta M}{(HT)^{2/3}}
\end{equation}
and
\begin{equation}
t_{\rm scal}\equiv \frac{T-T_c(H)}{(HT)^{2/3}},
\end{equation}
where $T_c(H)=T_c[1-H/H_{c2}^c(0)]$. 
This scaling was probed by using the above determined $T_c$ (see Table II), and $H_{c2}^c(0)$ as the only free parameter. The $H_{c2}^c(0)$ value that minimizes the $\chi^2$ with respect to a reference isofield data (4~T) is 45~T, although values between 40 and 50~T still lead to very similar scalings (the corresponding $\chi^2$ are within $\sim1$\%). As it may be seen in the main panel of Fig.~\ref{Figscal}, the 3D scaling is confirmed in spite of the noise affecting the largest applied magnetic fields. The associated in-plane coherence length amplitude is $\xi_{ab}(0)=27.0\pm1.5$~\r{A}.


The 3D behavior in the critical region may still be consistent with a 2D behavior well above $T_c$ if the transverse coherence length $\xi_c(T)$ shrinks to values well below the interlayer distance $s$. In fact, this is the case of a well known quasi-2D superconductor like optimally-doped YBa$_2$Cu$_3$O$_{7-\delta}$, which presents a 3D behavior in the critical region,\cite{Welp91} and a 3D-2D transition in the Gaussian region well above $T_c$, at reduced temperatures around 0.1.\cite{bookLarkin} 

\subsection{Analysis in the Gaussian region}

\begin{figure}[t]
\begin{center}
\includegraphics[scale=.5]{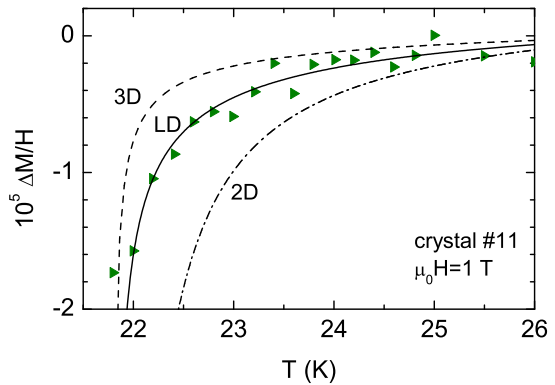}
\caption{(Color online) Temperature dependence just above $T_c$ of the fluctuation magnetic susceptibility obtained with $\mu_0H=1$~T. The solid line is the best fit of Eq.~(\ref{MLD}) above 22~K with $r$ as the only free parameter (see the main text for details). For comparison, the result for the 2D and 3D approaches are also included (see the main text for details).}
\label{FigflucM}
\end{center}
\end{figure}

The fluctuation magnetic susceptibility in the Gaussian region above $T_c$ is presented in Fig.~\ref{FigflucM}. This measurement correspond to $\mu_0H=1$~T. Lower applied magnetic fields lead to a proportionally lower signal to noise ratio, and for fields above 1~T the SQUID's sensitivity decreases significantly. In addition, 1~T is still much smaller than $\mu_0H_{c2}^c(0)\approx45$~T, so that the data are in the so-called low-field (or Schmidt) limit in which finite-field effects may be neglected. In this limit the LD model under a total-energy cutoff leads to\cite{Rey17}
\begin{equation}
\frac{\Delta M}{H}=-\frac{\pi\mu_0k_BT\xi_{ab}^2(0)}{3\phi_0^2s}\left[\frac{1}{\sqrt{\varepsilon(\varepsilon+r)}}-\frac{1}{\sqrt{c(c+r)}}\right].
\label{MLD}
\end{equation}
Here $k_B$ is the Boltzmann constant, $\mu_0$ the vacuum magnetic permeability, $\phi_0$ the flux quantum, and $c=\ln(T_{\rm onset}/T_c)$ the total-energy cutoff constant. When the LD parameter is $r\ll1$ or $r\gg1$, and in absence of cutoff ($c\to\infty$), this expression reduces to the classic results for the 2D or 3D limits, respectively. By using the $T_c$ and $\xi_{ab}(0)$ values determined in the analysis of the critical region, and $c=0.23$ as corresponds to the above determined $T_{\rm onset}=27.5$~K, the analysis depends only on $r$. The solid line in Fig.~\ref{FigflucM} is the best fit to the experimental data down to 22~K, which is very close to $T_c$. Below this temperature, the theory strongly overestimates the measured $\Delta M/H$ amplitude, which may be due to the onset of critical fluctuations. Note also that $T_c$ inhomogeneities are expected to play a non negligible role for temperatures above $T_c[1+\Delta T_c/T_c-H/H_{c2}^c(0)]$, which is close to 22~K. The LD parameter resulting from the fit is $r=0.13\pm 0.03$, where the uncertainty comes from the one in the $H_{c2}^c(0)$ value. The $r$ value is well below the onset reduced temperature, which confirms the quasi-2D nature of this material. The associated transverse coherence length is $\xi_c(0)=1.9\pm 0.2$~\r{A}, that when combined with the above $\xi_{ab}(0)$ leads to an anisotropy factor as high as $\gamma=\xi_{ab}(0)/\xi_c(0)=14\pm 1$. The superconducting parameters for crystal \#11 are also summarized in Table~II. Just for completeness, the dot-dashed and dashed lines in Fig.~\ref{FigflucM} are the 2D and 3D limits of Eq.~(\ref{MLD}), evaluated by using $r=0$ and 0.55, respectively (this last reference value corresponds to the $\xi_c(0)$ value of the 3D crystal \#9).

\section{Discussion}

\subsubsection{Bulk nature of the superconductivity}

The detailed characterization presented in Ref.~\onlinecite{Luo17} shows the bulk nature of the superconductivity in these compounds. The agreement of $\Delta M$ with the LD theoretical approach, and the fact that the resulting superconducting parameters are within the ones obtained from $\Delta\sigma$, further confirms this point, and also our conclusions about the high anisotropy and quasi-2D behavior of these materials. It is worth noting that $\Delta\sigma$ is also sensitive to the superconducting volume fraction; if it were small, $\Delta\sigma$ would be reduced roughly in the same proportion,\cite{Maza91,Rey17} and the analysis would be inconsistent. However, our results agree with the theoretical approaches by assuming a full superconducting volume fraction. Finally, the specific-heat jump at $T_c$ (also directly proportional to the superconducting volume fraction) has been measured in crystals of a similar composition (Co-doped instead of Ni doped).\cite{Jiang16_2} It is found $\Delta C_p/T_c=6.7$~mJ/(mol~Fe~K$^2$), that follows the $\Delta C_p/T_c$ vs. $T_c$ scaling reported in Ref.~\onlinecite{Zaanen09} for different IBS.

\subsubsection{$T_c$ dependence of the superconducting parameters}

\begin{figure}[t]
\begin{center}
\includegraphics[scale=.5]{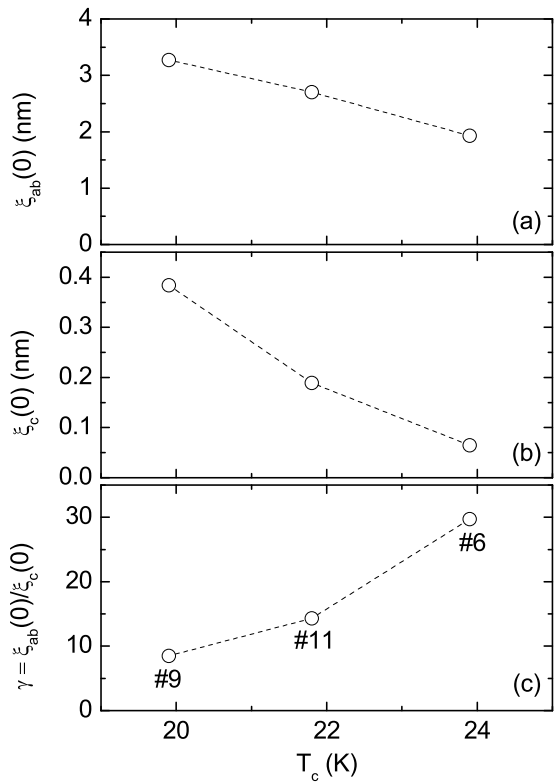}
\caption{$T_c$-dependence of the coherence lengths and of the anisotropy factor of the studied samples.}
\label{Figparam}
\end{center}
\end{figure}

As it may be seen in Fig.~\ref{Figparam}, the $\xi_{ab}(0)$ and $\xi_c(0)$ values obtained in the above analysis decrease as $T_c$ increases. The dependence is more pronounced in the case of $\xi_c(0)$, and as a consequence the anisotropy factor presents an steep increase with $T_c$, reaching a value as high as $\sim30$ in crystal \#6. This sample presents the highest $T_c$ and the smallest La-doping level (0.172) of the studied samples. So it may be concluded that $\gamma$ decreases for doping levels above the optimal one, that is expected to be about 0.15.\cite{Kudo14,Kawasaki15} Such a behavior is opposite to the one observed in Ni-doped 122 single crystals, for which $\gamma$ was found to increase upon increasing the Ni content above the optimal one.\cite{RamosSST15}  

The $\xi_c(0)$ reduction with the La doping level may be related to a weakening of the FeAs layers coupling. In fact, in the LD model of Josephson-coupled superconducting layers, the $c$-axis coherence length amplitude is related to the interlayer coupling constant $\Gamma$ through $\xi_c(0)=s\sqrt{\Gamma}$.\cite{Klemm75,*Ramallo96} Then, we may assume that the La doping strongly affects the Josephson coupling between the FeAs layers. This is consistent with recent results on the electronic structure of Ca$_{0.85}$La$_{0.15}$FeAs$_2$, where it is concluded that \textit{the Ca-La layers not only supply carriers but also tune the coupling between the As chains and the FeAs superconducting layers}.\cite{Liu15}

\subsubsection{Comparison with the anisotropy factors in other IBS}

To our knowledge, the $\gamma$ values found here (very in particular the one of crystal \#6) are among the largest ever reported in IBS. For instance, 1111 compounds present $\gamma(T_c)\approx6-9$,\cite{Jaroszynski08,Lee09,Karpinski09,Welp11} very recent works on FeSe intercalated with Li-NH$_3$ and with Li$_{1-x}$Fe$_x$OH reported $\gamma(T_c)\approx15$,\cite{Sun16,Yi16} and in highly overdoped Ba(Fe$_{1-x}$Ni$_x$)$_2$As$_2$ ($x=0.1$) it is found $\gamma(T_c)\approx16$. In a recent work on the anisotropic properties of a crystal from the 112 family (Ca$_{1-x}$La$_x$Fe$_{1-y}$Co$_y$As$_2$ with $x=0.2$ and $y=0.02$), it is reported $\gamma\sim5$ near $T_c$.\cite{Xing16} The difference with the much larger $\gamma$ values obtained in our crystals could be attributed to the smaller doping level. In fact, while the La concentration is similar, the Co concentration is half the one of Ni in our crystals, and also the Co valence is smaller than that of Ni. 

It is worth noting that the majority of works in the literature obtain the anisotropy factor and coherence lengths from the shift of the resistive transition with $H$. However, in the present case, the large resistivity rounding due to quasi-2D fluctuations would introduce a large uncertainty (the result would be strongly dependent on the criterion used). In turn, procedures based on the analysis of the angular dependence of the magnetoresistivity around $T_c$ in terms of the 3D-anisotropic GL approach,\cite{Yuan15,Xing16,Yi16} may not be applicable to quasi-2D superconductors.

\subsubsection{Quasi-2D behavior}

The $\xi_c(0)$ value resulted to be significantly smaller than the FeAs layers interdistance, $s=10.34$~\r{A}. In samples \#6 and \#11 this leads to a LD parameter well below the onset reduced temperature, so that a 3D-2D transition (a quasi-2D behavior) is observed at accessible reduced temperatures. \cite{[{It is interesting to mention the fact that materials with smaller $s$ values like BaFe$_{2-x}$(Ni,Co)$_x$As$_2$ may still present 2D spin excitations, and 3D (anisotropic) low-energy excitations within FeAs-plane. See, e.g., }]Lumsden09, *Luo13} As commented on in the Introduction, early works suggest a 2D behavior in compounds with even smaller $s$ values,\cite{Song12,Rullier12,Pandya10,Pallecchi09,Liu10_1,*Liu10_2} but these results are not confirmed in more recent works in the same or similar compounds.\cite{Putti10,Welp11,Liu11SSC,Pandya11,comment,Ahmad14,Ahmad16} A 3D-2D transition has been recently proposed in 10-3-8 single crystals (with $s=10.7$~\r{A}) in Ref.~\onlinecite{Ahmad16b}, after the observation of a change in the critical exponent of $\Delta\sigma(H=0)$ from $-1/2$ to $-1$. However, the $\xi_c(0)$ value found by these authors is close to $s/2$, which would be rather consistent with a 3D behavior up to high $\varepsilon\sim(2\xi_c(0)/s)^2\sim1$. It would be then interesting to check whether the seeming 2D critical exponent observed at $\varepsilon\approx 0.1$ in these compounds can be explained in terms of short-wavelength effects. 

The combination of a large anisotropy and a large FeAs interdistance makes accessible the field scale above which 2D vortices would appear in the mixed state $\sim\phi_0/s^2\gamma^2$ (see, e.g., Ref.~\onlinecite{tinkham}). In fact, in crystal \#6 this field would be only $\sim2$~T. This makes this compound a possible candidate to study 2D vortex physics in IBS.

\section{Conclusions}

We have presented detailed measurements of the conductivity and magnetization induced by superconducting fluctuations near $T_c$ of several high-quality single crystals of the 112 family, in particular Ca$_{1-x}$La$_x$Fe$_{1-y}$Ni$_y$As$_2$ with $x=0.17-0.20$ and $y=0.044(3)$. As compared to the more studied 11, 111, 122 and 1111 families, this compound presents an extra As-As chain spacer-layer that increases the FeAs layers interdistance up to $s=10.34$~\r{A}, and it is expected to be strongly anisotropic. The data were then analyzed in terms of a generalization of the Lawrence-Doniach model to finite applied magnetic fields and high reduced temperatures through the introduction of a total-energy cutoff. This allowed a precise determination of fundamental superconducting parameters like the in-plane and transverse coherence lengths. The resulting anisotropy factors are among the largest observed in IBS (up to $\sim30$ in the highest $T_c$ crystal), and are directly correlated with the $T_c$ value. This comes mainly from a significant $\xi_c(0)$ dependence on $T_c$, which may be related to a dependence of the interlayer coupling on the La-doping level. In the higher $T_c$ crystals $\xi_c(0)$ is much smaller than the FeAs layers interdistance, $s$, leading to a 2D behavior at accessible reduced temperatures. In spite of this, the non-vanishing LD parameter is still consistent with a non-negligible coupling between adjacent FeAs layers, and then between the FeAs layers and the As chains, which seems to be a requisite for the existence of topological superconductivity in these compounds. 

It would be interesting to extend our present results to a wider range of La- and Ni-doping levels, and to other IBS families with large FeAs interdistances, like 10-3-8 and 10-4-8 (also with intermediate As layers in the spacer layer),\cite{1038_1,1038_2,1038_3} 32522,\cite{Wen09_1} 42622,\cite{Ogino09,Wen09_2} (Fe$_2$As$_2$)[Ca$_{n+1}$(Sc,Ti)$_n$O$_y$] $(n=3,4,5)$,\cite{Ogino10} and 1144 (e.g. CaKFe$_4$As$_4$).\cite{Iyo16}

\begin{acknowledgments}

We thank Prof. F\'elix Vidal for valuable discussions and remarks on the manuscript. This work was supported by the Spanish MICINN, project FIS2016-79109-P (AEI/FEDER, UE), and the Xunta de Galicia (projects GPC2014/038 and AGRUP 2015/11). The work at IOP, CAS is supported by NSFC (projects: 11374011 and 11674406),  MOST of China (973 projects: 2012CB821400 and 2015CB921302), and the SPRP-B of CAS (Grant No. XDB07020300). H. Luo is grateful for the support from the Youth Innovation Promotion Association, CAS (No. 2016004).

\end{acknowledgments}

\bibliography{paper}

\end{document}